\documentclass[12pt]{article}
\usepackage[latin1]{inputenc}
\usepackage{amsmath,amsfonts,amssymb,amscd,amsthm,graphicx}
\DeclareGraphicsExtensions{.pdf,.png,.jpg}
\usepackage[pdfpagelabels,plainpages=false]{hyperref}

\begin{document}

\begin{abstract}
This is a survey on the One Time Pad (OTP) and its derivatives, from its origins to modern times. OTP, if used correctly, is (the only) cryptographic code that no computing power, present or future, can break. Naturally, the discussion shifts to the creation of long random sequences, starting from short ones, which can be easily shared. We could call it the Short Key Dream. Many problems inevitably arise, which affect many fields of computer science, mathematics and knowledge in general. This work presents a vast bibliography that includes fundamental classical works and current papers on randomness, pseudorandom number generators, compressibility, unpredictability and more.
\end{abstract}

	\section{The beginning of modern cryptography}
	There are several books on the history of cryptography, see \cite{kahn},   \cite{lewand}, \cite{flbauer},\cite{bauer}, 
	\cite{bauer2}, \cite{dooley}, \cite{klima}, \cite{megyesi}. These texts also contain the basics of cryptography. You can find both the history of cryptography and its modern addresses in William Easttom's beautiful book \cite{easttom}.
	
	We can trace the beginnings of modern cryptography back to 1467, the year of the publication of the book \textit{De Cifris}, by Leon Battista Alberti.
	In this text, an encryption tool is described in detail. There are two concentric discs, one of which is movable and can rotate inside the other. The symbols of the alphabet are imprinted on the discs, so that a rotation of a certain angle corresponds to a permutation of the alphabet itself.
	The method therefore consists in replacing one symbol with another, in a reversible way, as in the ancient so called monoalphabetic substitution codes.
	However, there is a fundamental difference: the permutation may be changed for each letter of the text. For this reason these are called polyalphabetic substitution codes, or Vigenère codes.  Blaise de Vigenère, in his book \textit{Traicté des Chiffres}, written in 1586, divulged and improved the ideas of Alberti and his successors Giovan Battista Bellaso and Giovanni Battista Porta. Vigenère's fundamental contribution can be seen in the explicit use of the key (see \cite{bauer}, Chapter 3).
	
	We describe the polyalphabetic substitution method with modern terminology.
	
	We call alphabet a set $\mathcal{A}$ of $q$ symbols.
	$\mathcal{S}$ is the set of all permutations of $\mathcal{A}$.
	A string of permutations 	$s=\sigma_1 \, \sigma_2 \dots \sigma_k$, where $\sigma_i \in \mathcal{S}$, is a key. We call $k$ the length of the key.
	A message $M$ is a string of symbols from 	$\mathcal{A}$, $M=x_1x_2\dots x_t$.
	
	$M$ is encoded by the function $E_s$ in this way
	\begin{equation} \label{vigenere}
	E_s(M)=\sigma_1(x_1) \, \sigma_2(x_2) \dots \sigma_k(x_k) \, \sigma_1(x_{k+1}) \, \sigma_2(x_{k+2}) \dots
	\end{equation}
	
	Vigenère's idea is very effective. The statistic of the message is completely destroyed. The statistic of the message is the frequency distribution of the $ q $ symbols that appear in the message. An investigation based 
	on the relative frequencies of letters is useless. However
	in 1863 Friedrich Kasiski realized that, if the keyword length $ k $ is known, the problem of breaking a
	polyalphabetic substitution code can be reduced to that of deciphering $ k $ monoalphabetic codes.
	
	In fact the symbols $ x_1, x_ {k + 1}, x_ {2k + 1}, \dots, x_ {hk + 1} $ will be encrypted by the permutation $ \sigma_1 $,
	the symbols $ x_2, x_ {k + 2}, x_ {2k + 2}, \dots, x_ {hk + 2} $ will be encrypted by the permutation $ \sigma_2 $, and so on.
	
	This is a truly \emph {algebraic} idea: a complex code is broken into the \emph{direct sum} of $ k $ simple codes.
		
If the keyword has length $ k $, we construct, taking one letter every $ k $, $ k $ messages, each one of which has been encoded with a single permutation $ \sigma_i $.

Finally a statistical attack is used on each of the $ k $ messages found (see e.g. \cite{bauer}, 3.3).

This attack is possible, because a one-letter substitution code does not change the statistic of the message.

Kasiski himself proposed a method for finding the length of the key, but today we have much more effective systems, for example the Friedman index.

Despite this weakness, we observe that there are still studies and applications of Vigenère's method (\cite{grosek}, \cite{ahamed}, \cite{park}, \cite{unyial}).

\section{The Friedman index}

William Friedman, an eclectic scientist (\cite{goldman}), was one of the most renowned cryptographers in history. He even studied (\cite{reeds}) the famous  Voynich Manuscript! He introduced the coincidence index in 1922 (\cite{friedman}). This is a very fundamental idea. Given a text T written by using $q$ different characters, the index of coincidence $I(T)$ is the probability that by taking at random two symbols in the text, they are equal. Supposing that $T$ contains $n$ characters and that the $i-th$ symbol appears $n_i$ times, then $I(T)$ is given by the formula
	\begin{equation}\label{indice}
I(T)=\sum_{i=1}^{q}\frac{n_i(n_i-1)}{n(n-1)}
\end{equation}
By calculating the average of $I(T)$ for many texts written in a given language \textbf{L}, we determine a coincidence index for \textbf{L} itself, $I$(\textbf{L}).
We call random language \textbf{R}($q$) that one in which each of the $q$ characters is randomly selected with probability $\frac{1}{q}$. Obviously $I$(\textbf{R}($q$)) = $\frac{1}{q}$.
If we encrypt a text $T$ with a Vigenère cipher, with key $K$, obtaining the ciphertext $E_K(T)$, we will observe that $I(E_K(T))$ approaches $I$(\textbf{R}($q$)) by increasing the length $k$ of the key. This fact can be used to determine the length of the key.
Let's take a couple of examples..

First of all, let's update the Vigenère cipher, so that we can encode every byte string.

Symbols are bytes, which are $8$-bit numbers, ranging from $0$ to $255$.
A text $T$ is a string of bytes of length $n$.
Key $K$ is a string of bytes of length $k$.
We then apply the method (\ref{vigenere}), where the permutations $\sigma_i$ are simply cyclic shifts of the byte sequence.
Coding starts by repeatedly writing $K$ under $T$ and then adding byte
per byte modulo $256$. For example, if T = $(144, 90, 69, 102)$, $k=3$ and $K=(70, 200, 240)$ then the text $T$ encrypted with the key $K$ is
$$E_K(T)=(144, 90, 69, 102)+(70,200,240,70) \mod{256}=$$ $$(214, 34, 53, 172)$$

So from our point of view, a text is simply a finite sequence of bytes. Any file on our computer can be considered text, reading it one byte at a time. Notice that we have $256$ characters, but each text has its own particular $q$, which is the number of distinct bytes that appear in it. Given a text $T$ and an integer $d$, we define $T_d$ as the text that is obtained from $T$ taking one character every $d$. Finally, we define $I(T,d)$ as the function $I(T,d)= I(T_d)$.

Suppose $T$ was encrypted with a key $K$ of length $k$ and $C_K= E_K (T) $. Then all the characters of $ C_K $ have been encrypted with the same permutation and, in the graph of $ I (C_K, d) $, as $ d $ varies, we will observe peaks corresponding to the multiples of $ k $. We will then find the length of the key.

Let's come to the examples. The $ T $ text considered is the dante.txt file, the complete text of the Divine Comedy. It contains $ 573753 $ characters.

We calculate the coincidence index of $ T $ with the formula (\ref{indice}). In this case $ q = 82 $, because there are $ 82 $ different bytes.

We get $ I (T) = $ 0.06217, while the index of a random text with $ 82 $ symbols is $ 1/82 = $ 0.01219.

The graph of the function $ I (T, d) $, with $ 1 \leq d \leq 200 $ is

\begin{center}
	\includegraphics[scale=0.5]{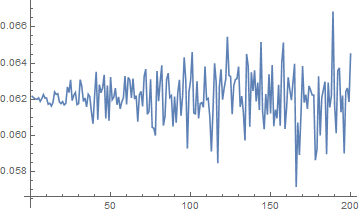}
\end{center}

Now we encode $ T $ with Vigenère and a random key $ K $ with $k=30$.

We have $ C = E_K (T) $.

The graph of the function $ I (C, d) $, with $ 1 \leq d \leq 200 $ is

\begin{center}
	\includegraphics[scale=0.5]{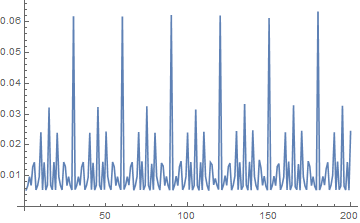}
\end{center}	
	
In $ C $ there are $ q = 256 $ distinct characters (i.e. there are all), $ I (C) = 0.00590 $ and $ 1/256 = 0.00390 $.

It is clear that the highest peaks are at multiples of $ 30 $.
The minor peaks correspond to the length divisors. In this way, by examining only the ciphertext $ C $, we find out the length of the key!

Note that this technique does not require you to know in which language the text is written, it does not make direct use of the $ I$(\textbf{L})  index. It is based solely on the fact that the language used is structured, not casual. That is, it is based on the difference between $ I$(\textbf{L})  and $ I$(\textbf{R}($q$)) . This difference is seen in the peaks of the graph, which appear when $ d $ is a multiple of the key length. In fact, in this case, the sub-text examined has been encrypted with a mono-alphabetical substitution code, and therefore it maintains the statisics of the  language  \textbf{L} .
	
So let's take an example with a completely different type of file, an exe file. Let's take the gp.exe file, the executable of the beautiful computer algebra system  \href{https://pari.math.u-bordeaux.fr/download.html}{Pari/Gp}.

This is the graph of $I(gp.exe,d)$, $1 \leq d \leq 200$

\begin{center}
	\includegraphics[scale=0.5]{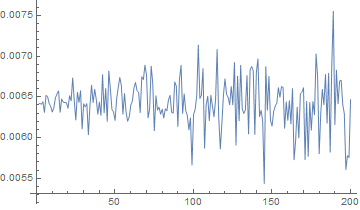}
\end{center}

In gp.exe contains $245248$ bytes, and $q=256$. 

We have $I(gp.exe)=0.00640725$, while $I$(\textbf{R(256)})$=0.00390625$.

As before we encode $ gp.exe $ with Vigenère and a random key $ K $ with $k=30$.

We pose $ S = E_K (gp.exe) $.

The graph of the function $ I (S, d) $, with $ 1 \leq d \leq 200 $ is

\begin{center}
	\includegraphics[scale=0.5]{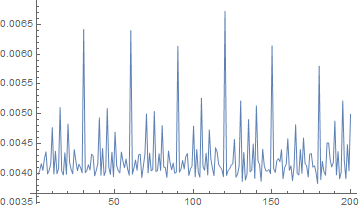}
\end{center}

The index of the encoded text $S$, $I(S), $dropped from the original $0.00640725$ to  $0.00397486$. Peaks are clearly visible at multiples of $30$.

Using the Friedman index, by analyzing the graph of $ I (S, d) $, whoever intercepts the encrypted message $ S $ can find out if it has been encoded with a polyalphabetic substitution code, and know the length of the key, without knowing anything about the nature of the original text. Obviously, not knowing the language of the source, it will not be possible, in the decryption phase, to use a statistical attack.

As noted in \cite{alkazaz}, although substitution codes are not safe in themselves, they are used as constituent parts of other, more powerful codes. It is therefore important to thoroughly examine the vulnerability of these systems. And it is necessary to do it automatically, given the immense amount of data in circulation. Several types of algorithms are used, including genetic and compression, see \cite{dureha}, \cite{alkazaz}, \cite{hilton}, \cite{sabonchi}.

A variant of the Friedman index, called the Progress Index of Coincidence (PIC), was used in \cite{sommervol} to define a good fitness function for a genetic algorithm that is capable of breaking the Enigma code.

The index of coincidence is also used in fields other than cryptography. For example in \cite{vera}, where a distance between human languages is introduced, and in \cite{gagnuc}, where the use of this index is proposed to determine
new patterns and evolutionary signatures in DNA sequences.

The Friedman index (also known as the Kappa index) is an important research tool, and certainly deserves further study on it.

\section{Enigmatic Perfection}
In polyalphabetic substitution ciphers, the key must be long. The longer the key, the closer the Friedman index of the text approaches the index of the random language, and, consequently, the peaks in the graph are hardly detectable.

\noindent
Furthermore, of course, the key must be unpredictable.

In Germany, a machine was produced, the Enigma machine, which transformed a plaintext into the ciphertext, changing the permutation used for each letter that was written.
If the source text is formed by the characters $x_1,x_2,\dots,x_n$ then the $x_i$  is encoded with the permutation $\sigma_i$. The permutations are changed by rotating some disks.
The permutations were repeated periodically, but the period (i.e. the length of the keyword) was very large, tens of thousands of characters.
The Enigma code was broken by the collective work of two teams of researchers, resident in England and Poland. The best known of them is Alan Turing (\cite{turing}), one of the founders of computer science and complexity theory.

For the history and operation of the Enigma machine see \cite{bauer} Ch. $8$, \cite{dooley} Ch. $8-9$, \cite{klima} Ch. $4-5$, \cite{christensen},  \cite{thimbleby}, \cite{kenyon}.

The events that marked the breaking of the Enigma code were really interesting, even at a theoretical level and developed, among other things, a fruitful dispute between algebraists and probabilists, see \cite{megyesi} p. $70-76$.

Trying to decrypt some short messages, encoded with the Enigma machine, is still a challenge today, and can be very instructive, see \cite{schrodel}, \cite{ostwald}, \cite{ostwald2}.

Tackling the Enigma remains an excellent benchmark for new automatic decryption methods.

To proceed further, it is useful to reduce the Vigenère method to the essential. Let us consider the alphabet formed only by the two symbols $ 0 $ and $ 1 $, the bits. A key is a string of permutations of the alphabet. In our case there are only two permutations: the identical one and the one that swaps $ 0 $ and $ 1 $. We can then forget about permutations and instead use the bit XOR, that is the sum modulo $ 2 $. The key becomes a string of bits, $ 0 $ stands for the identical permutation and $ 1 $ stands for the swap.
We want the key to be long, so take it as long as the message! We want it to be unpredictable,so  we choose the bits at random!

This is the OTP (One-Time Pad), the central figure of our story.

In \cite{bellovin} it is proved that the inventor of the OTP was Miller, who discovered it about $ 35 $ years before the two re-discoverers, Vernam and Mauborgne. Usually the OTP is called Vernam code or perfect code.

We recall the definition of the OTP code.

The  message $ M $ is a $ n $ bit string, of arbitrary length $ n $.
The  key $ K $ is also a $ n $ bit string.
$$ M = x_1, x_2, \dots, x_i, \dots, x_n $$
$$ k = k_1, k_2, \dots, k_i, \dots, k_n $$
The encrypted massage $C$ is obtained by XORing the bits of the message with those of the key.
$$ C = c_1, c_2 ,, \dots, c_i, \dots, c_n $$
where $ \forall i, \, c_i = x_i + k_i \mod 2 $.

Four conditions must be met for the OTP code:
\begin{enumerate}
	\item the key must be as long as the message\label{itm:1}
	\item the key must be random\label{itm:2}
	\item the key must be used only once\label{itm:3}
	\item the key must be kept secret.\label{itm:4}
\end{enumerate}

In $ 1949 $ Claude Shannon (\cite{shannon}) proved that the OTP code is (the only) perfect code.
This means that if the text $T$ has been correctly encrypted with OTP, obtaining the text in cipher $C$, whoever intercepts $C$ cannot obtain any information about $T$, regardless of the computing power at his disposal.

The reason is this: $C$ can come with the same probability from any text $U$, of the same length as $T$.

Even the brute force attack is not effective. If $T$ is long $ n $ there are $2^n$ texts of equal length. And there are $2^n$ keys $ K $. If we take all the keys one by one and calculate $C$ XOR $K$ we will find, in an unpredictable order, all the possible messages $U$, and we will never know which one was really sent!

It is clear that there is a major key management problem in the OTP. The key must not only possess the qualities \ref{itm:1},\ref{itm:2},\ref{itm:3},\ref{itm:4}, but, of course, it must be shared between those who send the message and those who can legitimately receive it. Now the question arises, if $A$ and $B$ can share, on a secure channel, a $ n $ bit key, why don't they directly share the message $M$, which has the same length?

There are  many ways to overcome these difficulties. OTP has actually been used in the past, in communications that required a very high degree of security (\cite{bauer} p. $103-115$, \cite{kahn} p. $714-731$). 

Particularly interesting was the SIGSALY system, designed to encrypt telephone conversations.

With the Sigsaly machine an attempt was made to realize OTP in the transmission of voice conversations. The voice was digitized and compressed (to save bandwidth). Eventually it was represented by strings of integers between $ 0 $ and $ 5 $. A key string was added to the entry string modulo $ 6 $. The key was simply derived from a $16$ 
inch vinyl record. Obviously, the record had to be the same for whoever spoke and who received. In the decoding phase the key was subtracted modulo $ 6 $.

SIGSALY was never broken. Turing also took an interest in it and carried out his own project to improve it (\cite{floyd}, Ch. $7$).

SIGSALY weighed $50$ tons, and the record was enough for ten minutes of broadcast. Nowadays everything has changed, thanks to technology.

 We can easily use our laptop to transmit OTP encrypted texts, imitating the SIGSALY process. My message $M$ is a string of $n$ bits. The only problem is sharing a string of $n$ bits (the key) with the receiver $R$. There are billions of terabytes available on the net! $R$ and I download the same text from the network (or piece of music, or painting or any digital object), we perform the same preprocessing operation on it, to destroy any internal structures (for example we zip the file), we take $n$ bits from a certain position, and we have the common key $K$! So I send on the network $C$ = $M$ XOR $K$, and $R$, when it receives $C$, computes $C$ XOR $K$.

I could also transmit to $R$ the key $K$, which I produced, hiding it, by means of steganography, in an image file, or the like. We can share the address of a huge database with an algorithm that compute the sequence of the objects to be used and the starting points. This would provide a substantially unlimited key, to be used as many times as desired, always taking different parts. It is a pattern whose technical details can be bridged in many different ways. This is why there are so many articles that revolve around similar concepts.

It is not possible to quote all relevant articles, to trace a complete history of what we might call the actualization of the OTP. See these works and their bibliographies: \cite{ruby}, \cite{nagaraj}, \cite{david}, \cite{vobornik}, \cite{omolara}.

There are very recent applications to new cryptographic methods and modern technologies, for example to Single Sign-On (\cite{kihara}) and Wireless Communications (\cite{lig}).

A very interesting idea is to put together OTP and DNA. For the fundamentals of the theory see  \cite{gahani} and \cite{borda2}.

As observed in (\cite{zhangy}) there are at least two ways to use DNA: manipulate it directly in a laboratory with biochemical tools, or simply consider it as a code. The genetic code is a four base sequence A - adenine, C - cytosine, G - guanine, T - thymine, which can be easily converted into a binary sequence, by means of substitutions
$$ A \rightarrow 00 \quad  C \rightarrow 01 \quad G \rightarrow 10 \quad  T \rightarrow 11  $$ 
These binary strings are manipulated so that they can be used as keys in the OTP  (\cite{borda}).

A DNA based (biochemical)  method for random key generation and OTP management is presented in \cite{zhangy}. 
In \cite{peng}  a one-time-pad cipher algorithm based on confusion mapping and DNA storage technology is proposed. 
There is also a very recent implementation in Python (\cite{abdelghany}).

\section{OTP approximations}

One of the advantages of DNA encryption is that you can share keys using huge databases that are public (e.g. \href{https://www.ncbi.nlm.nih.gov/}{The National Center for Biotechnology Information}).

However, in reality, as we have already noted above, we can try to use OTP easily. It is sufficient (for example) that one of us gives or sends to the other one a DVD containing a long string of bits ($\approx 10^{14}$). The string, with a particular segmentation and synchronization program, can be used as a key for many encrypted OTP communications.

Can we trust the keys taken from our disk? Let's look at the question.
Shannon (\cite{shannon}) showed that to achieve perfection it is necessary that the space of the keys be as large as that of the messages. Therefore $2^n$ keys must be available. That is, any binary string of length $n$ must be available as a key.
The single key used can be any string. What matters is that we must have chosen it, with uniform probability, in a basket that contains all the strings of length $n$.
It is therefore not very credible that the conditions necessary for the application of Shannon's Theorem are verified.

Our dream would be this: to be able to use a short key, which can be easily shared on a secure channel, and then use it with OTP in its perfection.
It seems an impossible dream, but it can be realized if we are satisfied with an approximate perfection.

In a seminal paper from $1992$, Ueli M. Maurer presented a new approach in which a public source of sequences of random bits is used. Let's see in detail Maurer's method.

Suppose that user $A$ wants to send $B$ the $n$-bit message M,  $M=(x_1,x_2,\dots,x_n)$.

The publicly-accessible $R$ is an array of independent and completely random binary random variables. 
$R$ consists of $m$ blocks of length $T$.
The block $i$, with $1 \leq i \leq m$ contains the bits $(R[i,0]$, $R [i,1]$, $\dots$, $R[i,T-1])$.

Now $A$ creates a secret key $Z$, $Z=(Z_1,Z_2,\dots,Z_m)$, where $0 \leq Z_h \leq T-1$ for every $h$. $Z$ must be chosen from the set $\left(0,1,\dots,T-1\right)^m $ with uniform probability. $A$ sends $Z$ to $B$ on a secure channel.

Using $Z$, $A$ builds the key $K$, to be used with OTP.

We set $S[h]=(R[h,Z_h+j-1]  \mod T)$ with $1 \leq h \leq m$ and $j=1,\dots,n$.

Finally the OTP key is $K=\oplus_{h=1}^{m} S[h]$.

$A$ sends to $B$ the message $C=K \oplus M$. Even $B$ can calculate $K$, because he knows $Z$ and $R$ is public, and finds $M=K\oplus C$.

We always assume that $T \gg n$. $R$ contains $L = mT$ random bits. The binary length of the secret key $Z$ is $ \approx m \log_2 T$. Those who know $Z$ must examine only $mn$ of the $L$ bits, that is a very small fraction $n/T$. Let's imagine that the opponent can, with the best strategy (even probabilistic), examine $N$ bits.
Mauer proves that if a certain event $\mathcal{E}$ occurs, the code is safe in Shannon's sense, i.e. the opponent cannot obtain any information about the plaintext $X$ from the ciphertext $C$.
The nature of the event $\mathcal{E}$ is not important, what matters is its probability, $P(\mathcal{E})$. Maurer proves that $P(\mathcal{E}) = 1-n \delta^m$.
This probability is extremely high, because $\delta=\frac{N}{L}$.

Maurer's Theorem is based on the fact that the opponent is storage-bounded, and can only examine a delta fraction of the bits of $R$.

Let's take an example. I want to send $B$ a document $X$ containing $2^{27}$ bits, about $100$ Mb. With the classic OTP I would have to share with $B$ a random string (the key) of $100$ Mb. Instead, I apply Maurer's method assuming $m = 40$ and $T = 10^{10}$. Suppose that opponent's limit forces him to examine no more than $1/3$ of $R$, i.e. delta = $1/3$. I create a secret key $Z$, which will be about $1328$ bits long, as $m \log_2(T) \approx 1328$. I share $Z$ with $B$, and send the encrypted text $C$. By intercepting $C$, the opponent cannot have any information about $X$, with probability $1- 2^{27} (1/3)^{40} = 1-10^{-11}$.

Mauer's idea was revived, modified and perfected by Rabin and others, see \cite{ding}, \cite{rabin},  \cite{sadjadpour}.

In \cite{sadjadpour} the short key dream is essentially fulfilled. The final key, which complies with Shannon's requirements, is created through an ingenious process of manufacturing intermediate keys. With an iterative method, the author shows that the relationship between key length and data can be made as small as desired, at the cost of increased computational complexity. So, surprisingly, the birth of quantum computers or other extremely powerful devices will not facilitate the breaking of the code (which is impossible, as we know) but will make the encoding of messages with OTP very fast, and the key very short!

\section{The Need for Randomness}

We need randomness in everyday life. Pizza or pasta tonight? We flip a coin, or do odds and evens. As Hayes says in \cite{hayes}, there is a real Randomness Industry. Inside each slot machine there is a special chip that continuously calculates random numbers. Immense amounts of random numbers are used every second around the world in video games, simulations, optimization algorithms, probabilistic algorithms, Monte Carlo methods and, we know, cryptography.

Any cryptographic system requires the use of keys. A sure rule is that there are no secure systems with keys that are too short, for the simple reason that, if the key is $n$ bits long, with a brute force attack it is enough to find all $2^n$ keys. On the other hand, with a key of a few hundred bits we would like to encrypt messages of many thousands of bits.

We limit ourselves in this survey to stream codes, direct emanations of our OTP, which can be considered their prototype. 

It is important to note that many block ciphers, for example AES, can be used as stream codes, using techniques such as Output Feedback (OFB) and others, recommended by NIST (\cite{dworkin}, \cite{dworkin2}).

In general, the encoding in a stream code occurs exactly as in OTP. We have a stream of binary messages $m_1,..,m_t,..$ and keys $k_1,..,k_t..$, we get a stream of encrypted messages $c_j = m_j\oplus k_j$.

The key stream is generated from an initial secret and, of course, random key $K$.
If $A$ and B want to communicate, they share $K$ and both generate the same key stream. For reference texts see \cite{rueppel} and \cite{klein}.

During the $10$th International Conference on the Theory and Application of Cryptology and Information Security ($2004$), Adi Shamir gave a lecture entitled \textit{Stream Ciphers: Dead or Alive?} (\cite{shamir}). In his presentation, Shamir talked about a decline of stream ciphers, unlikely to be reversed in the near future. However, the author highlighted two particular areas, in which stream ciphers could have maintained priority. He said \textquotedblleft I believe that stream ciphers will remain competitive in two types of applications: 
a) hardware oriented scheme with exceptionally small footprint (gates, power consumption, etc)
b) software oriented scheme with exceptionally high speed\textquotedblright

These considerations sparked a lively discussion and in the same year ($2004$) \textit{eSTREAM: the ECRYPT Stream Cipher Project} was launched. The competition ended in $2008$. A full description of the finalists can be found in this book \cite{robshaw}. For a survey on stream ciphers see \cite{jiao}.

The research focuses on particularly fast or light stream codes.

Among the fastest codes there are Rabbit (\cite{boesgaard}) and Salsa family (\cite{bernstein}, \cite{muhammad}).

Lightweight Cryptographic Algorithms are increasingly important in the IoT. They are especially needed when dealing with small medical implants, battery-powered handheld devices, embedded systems, RFID and Sensor Networks, see Nist Internal Report \cite{mckey}. A detailed study of Low Energy Stream Ciphers is here \cite{banik}.

Everything we have seen requires the use of random bit strings.

\section{Randomness}

Let us begin by remarking that these bit (or number) sequences  are produced by programs running on ordinary computers, and are therefore completely deterministic.

Calling them \textit{random} seems strange. In fact they are said \textit{pseudo random}. Their generators are called PRNG, \textit{Pseudo Random Number Generators}.

Thousands of articles have been and will be written about them (\cite{beebe}).

What do we mean by a random sequence?

Goldreich, in \cite{goldreich}, summarizes the basic concepts of the three main theories very well.
\begin{enumerate}
	\item In his \textit{Information Theor}y ($1948$) Shannon characterizes perfect randomness as the extreme case in which the string of symbols does not contain any redundancy, i.e. there is a maximum amount of information.
	\item Solomonov ($1960$), Kolmogorov ($1963$) and Chaitin ($1965$), founded the second, computational theory. The complexity of a string is essentially the length of the smallest program that can generate it. In essence, if a string is truly random, a program must contain it in order to express it.
	\item  Blum, Goldwasser, Micali and Yao began, in the years $1982-84$, the third theory which pays attention to the actual computation. A sequence is random if we do not have computational procedures to distinguish it from a uniform distribution.
\end{enumerate}

	Chaitin-Kolmogorov's theory is fascinating, because it makes it possible to deal with the randomness of a single string, without resorting to any probability distribution (\cite{chaitin}, \cite{calude}, \cite{li}, \cite{downey}, \cite{shen}, \cite{franklin}).
	
	According to it we see that three concepts are essentially equivalent: randomness, incompressibility and unpredictability.
	
	This is also in accordance with our intuition. A random event cannot be predictable, and in order to compress a string of bits it needs to have some regularity.
	
	There are many efficient compression algorithms available today. The compressed text is expected to approach a random text. In this context, some compressors are studied in \cite{kleinst}, and it is proved that arithmetic coding seems
	to produce perfectly random output.
	
	If we delve into the subject, several surprises await us.

\section{Incompressibility}

We say that a string of $n$ bits is $c$-incompressible if it cannot be compressed more than $c$ bits. A simple counting argument (\cite{li}, p. $117$) shows that there are at least $2^n-2^{n-c}+1$ c-incompressible strings. So there is at least one $n$-bit string that cannot be compressed even by one bit, at least half of the strings are $1$-incompressible, the three quarts are $2$-incompressible and so on.

	The extreme majority of strings are very little compressible, and therefore highly random!
	
	Can any relationship exist between the infinity of prime numbers and the incompressibility of information?
	
	There are many different proofs of the infinity of primes. Often they are based on the fact that if the primes were finite, something would happen which is false. We have also made one \cite{cerruti2}. If primes were finite, N would be a  field!
	
	A proof I love is due to Chaitin  (see \cite{calude} p. 361). If primes were finite, almost everything would be compressible.
	
	Suppose there are only $k$ primes, $p_1,p_2,...,p_k$. Given an integer $N>1$, by the unique factorization theorem, we will have
	$$N=p_1^{e_1} p_2^{e_2} \cdots p_k^{e_k}$$
	Clearly we have $k \leq \log_2N$ and $\forall i, \, e_i \leq \log_2 N$.
	\medskip
	
	The number $N$ is identified precisely by the string of exponents $e_1,...,e_k$. These exponents are integers $\leq \log_2 N$ and can therefore be expressed (each) by $\log_2 \log_2 N$ bits.

	In conclusion, every integer $N$ can be expressed by
	\begin{equation} \label{numerobit}
	k\log_2 \log_2 N
	\end{equation}
	bits.
	\medskip
	
	Let us take a string $M$ containing $m$ bits. This uniquely identifies an integer $N$ of $m$ bits, of order $2^m$.
	\medskip 
	
	By (\ref*{numerobit}) $N$, and hence $M$, is determined by a string of $k \log_2 m$ bits.
	\medskip
	
	There exists $m_0$ such that 
	$$ \forall m>m_0 \quad m>k \log_2 m$$
	
	From this we deduce that \textit{all sufficiently long bit strings are compressible}!
	
	But we know very well that this is not true. 
	
	What if we insist on compressing everything?
	
	Of course this is possible, if we accept that we cannot go back, that we lose information.
	For example, given a message M of $n$ bits, with $n$  large, we can decide to take the first $128$ bits of M. This does not seem very useful.
	
	 However, it would be convenient to create, for each message M of length $n$, a kind of $128$-bit fingerprint, which would uniquely identify it. This is clearly impossible, if $n>128$, but hashing techniques try to come close to this dream.
	 
	Hash functions are very important in cryptography. A hash function $H$ transforms a message M of any length $n$ into a string of fixed length $m$. Typically $m$ is $128$ or $256$. 
	
	A cryptographic hash function $H$ must satisfy two conditions:
	
	- given $y=H(x)$ it must be computationally difficult to find $x$
	
	- it must be computationally difficult to determine $x$ and $x'$ such that $H(x) = H(x')$
	
	Two other properties, not easily formalized, are required in practice. 
	
	The hash $H(x)$ must appear random (here we go again) even if $H$ is perfectly deterministic. It is also required that $H$ is sensitive to initial conditions. In the sense that, if we change a single bit in $x$, about half of the bits of $H(x)$ change.
	
	Hash functions are public and intensively used, see Ch. $9$ of \cite{easttom}, Ch. $7$ of \cite{hwang}.
	
	Hash functions have a thousand uses, ranging from database indexing to electronic signature (it is much faster to sign $H(M)$ than M). They are used in many cryptographic protocols, such as bit commitment and password management.
	Precisely in this last area we have patented a system that contains a rather interesting hash function,based on the Chinese Remainder Theorem (\cite{cerruti4}).

	As is known, bitcoin was introduced by Satoshi Nakamoto in $2008$ (\cite{nakamoto}). 
	Hash functions are the real engine of the so-called mining that bitcoin uses: one must find an $x$ such that $H(x) \leq t$, where $t$ is a target $256$ bits string, see Ch. $2$ of \cite{dhillon} and $10.5$ of \cite{hwang}. Thanks to bitcoins, through hash functions, randomness passes directly into the economy, and earning capital becomes a worldwide lottery. Bitcoin mining involves an enormous consumption of energy. For an in-depth analysis of the impact of bitcoins on the economy and the environment, see \cite{badea}.
	
	\section{Unpredictability}
	
	We say that a sequence of bits $S=(b_n)$ is \textit{unpredictable} if it passes the next bit test (\cite{kneusel}, §$6.1$), i.e. if there is no polynomial time algorithm which, receiving the first $n-1$ bits of $S$ as input, returns the $n$th bit with probability $p> 1/2$. 
	
	In a truly masterful article (\cite{blum}) L. Blum, M.Blum and M. Shub examined the predictability of two PRNG.
	
	One of them is the \textit{quadratic generator}.
	
	Let $N = pq$, where $p, q$ are primes congruent to $3$ modulo $4$. We choose an integer $ x_0 $, the \textit {seed}.
	
	Then we produce the sequence
	$$x_{n+1}=x_n^2 \mod N$$
	The random bit sequence generated is $b_n=x_n \mod 2$.
	
	The authors prove that the sequence $ b_n $ is unpredictable, as long as you don't know how to factor $ N $.
	
	In their article $ BBS $ studied a second generator, besides the quadratic one, the $ 1/P $ generator, which is truly remarkable.
	
	$ P $ is a prime number, and $ b $ is a basis. If $ b $ generates the multiplicative group  $\mathbb{Z}_P^{*}$, the expression of $ 1/P $ in base $ b $ has period $ P-1 $. We can thus obtain very large periods, since today it is easy to find prime numbers with hundreds of digits.

	If $ b $ generates $\mathbb{Z}_P^{*}$, you get long strings of $b-$digits, which pass the statistical tests with good results.

	Surprisingly $ BBS $ show that the $1/P$ generator  is easily predictable: in fact it is sufficient to know $ 2l $ consecutive digits (where $ l $ is the length of $ P $ in the base $ b $) to deduce the entire period. The method is based on continued fractions.
	
	In general, the length of the period alone does not guarantee security. For example the \textit{Mersenne Twister} has a huge period, $2^{19937} -1$ but, as stated by the authors themselves (see (\cite{matsumoto})), it is not suitable for cryptographic purposes: indeed  the output of the algorithm becomes a linear
	recurring sequence by applying  a simple linear transformation.
	
	In prediction problems, it is natural to use Soft Computing or Artificial Intelligence techniques.
	
	A few years ago, with Mario Giacobini and Pierre Liardet (\cite{cerruti1}), we studied the prediction capabilities of Finite State Automata (FSA), evolving populations of automata using Genetic Algorithms.
	
	The underlying idea was to use the evolutive ability of prediction of the algorithm to get measures of the randomness of the sequence. Among other things we found that
	
	- the evolved prediction skills are in inverse proportion to the period length
	of the considered sequence;
	
	- the evolution of FSAs prediction skills seems to be directly linked to the
	linear complexity of the sequences considered.
	
	These last two conclusions make us hope that the evolution of FSAs could
	be used as a measure of the randomness character of a binary string.
	
	In (\cite{smith}) T. Smith considers automata of different types for the prediction of infinite strings with various types of periodicity.
	
	These machines, called predictors, can have multiple heads that at each step move along the assigned infinite sequence, read the symbol, change state and make a guess about the next symbol before it appears. This interesting paper describes new prediction algorithms for the classes of purely periodic, ultimately periodic, and multilinear words.
	
	The sequence of the digits of Pi (in any basis) is believed to be random (see §10). Therefore, the results of Fa and Wang are surprising. In (\cite{fenglei}) Fenglei Fa and Ge Wang use Neural Networks to make predictions on the bits of  Pi. Their neural networks predict the next bit (of $ 6 $ bit strings) with probability $>1/2$.
	
	The authors also apply the method to strings generated by PRNG. They conclude that neural networks, even very simple ones, can extract useful information for prediction from the data (if we are not in the presence of maximum entropy, or total disorder).
	
	In the remarkable article by Taketa et al. \cite{taketa} it is observed that neural networks learn to predict the next bit with a particular sensitivity to the linear complexity of the sequence.
	
	The field of machine learning in pseudorandom sequence prediction is just beginning and is truly fascinating.
	
	\section{Pseudo Random Number Generators}
	
	Most of us would like to have a TRNG (Truly Random Number Generator) available. On the net there are several sources, see \href{https://www.random.org/}{www.random.org}. 
	A now classic is \href{https://www.fourmilab.ch/hotbits/}{www.fourmilab.ch/hotbits}: HotBits are generated by timing successive pairs of radioactive decays detected by a Geiger-M\"{u}ller tube interfaced to a computer.
	
	This may not necessarily be the best choice. As Donald Knuth observes in \cite{knuth} p. $145$, "a truly random sequence will exhibit local nonrandomness"; there will certainly be, for example, sequences consisting of a million consecutive  zeros. OTP would send a million bits as plain text!
	
	Here, as we said, we make do with PRNGs.
	
	At the beginning of the chapter on random numbers (now a classic of the subject, recommended to all), Knuth recalls a fact from his youth. He built a very complicated algorithm to generate random sequences. Unfortunately when it was activated on the computer, sequences were observed that were repeated with a very short period! This is Knuth's (\cite{knuth}, p.$5$) conclusion:
	
	\textquotedblleft The moral of this story is that random numbers should not be generated with a method chosen at random. Some theory should be used. \textquotedblright
	
	Theory, examples and applications can be found here: \cite{easttom} Ch. $12$, \cite{johnston} and \cite{kneusel}.
	
	\cite{johnston} is particularly suitable for engineers and programmers, and also contains valuable information on the actual use of many programs.
	
	Also in \cite{kneusel} there is a particular attention to programming. It is a very well written text, rich in content. I found enjoyable the use of the  Monty Hall Dilemma (\cite{mlodinow}, Ch.$3$) which is proposed, with $C$-code and concrete examples, as a test of randomness!

	Cryptography requires random sequences of a particular type: see the overview \cite{marton} which states that one of the most important characteristics is that of unpredictability. In the survey \cite{aljahdali}, in addition to these problems, quasi-random numbers are also considered.
	
	In practice it is not possible to know a priori the qualities of a PRNG, it must be subjected to batteries of randomness tests. Many of them are currently in use, we quote Ent, Diehard, NIST, TestU01. See \cite{kneusel} Ch. $4$, \cite{johnston} Ch. $8-11$, the important article of Shen \cite{shen2} and the recent ACM  recommendations \cite{luengo}.

\section{Irrational Randomness}

As we saw in §4, talking about DNA and the ideas of Mauer and others, OTP can be realized starting from a large and public database of random bits (or numbers). If really big, we wouldn't even need to change it. Users only need to share the access points and formatting algorithms.
This is a great idea, and it is revived periodically, in ever new ways.

In \cite{coluccia} the author wonders if it would be better to extract rather than expand. The classical OTP is revisited, then the  expansion paradigm is compared to the idea of extraction. Users A and B only need to share a "mother pad". Gualtieri in \cite{gualtieri} suggests using the sequence of the digits of Pi. Naturally then each author proposes his own method to manage the common source of randomness.

But are Pi's digits really random?
Marsaglia's authoritative opinion is positive, see \cite{marsaglia} and \cite{marsaglia2}. There has been much debate on the question, see for example \cite{ganz}, \cite{bailey} and \cite{ganz2}. Empirically $\pi$ seems to pass all tests of randomness (\cite{mitsui}).

We try to understand how random $\pi$ is also in a qualitative way, mainly visual, look at the beautiful \textit{Walking on real numbers} (\cite{artacho}). $\pi$ is indistinguishable from the sequences generated by PRNG even in the fractals produced by Chaos Games (\cite{salau}). It seems that the first to walk on the $\pi$ digits was Venn (\cite{verbugt}). Pi was developed in base $8$, and each digit $0,..,7$ was associated with a direction. Of course numbers other than $\pi$ can be used, we have uncountable choices! Venn's idea is used below to see the first $4000$ digits of $\pi$, $e$ and $\cos(1)$ (left to right)

\begin{center}
	{\includegraphics[scale=0.3]{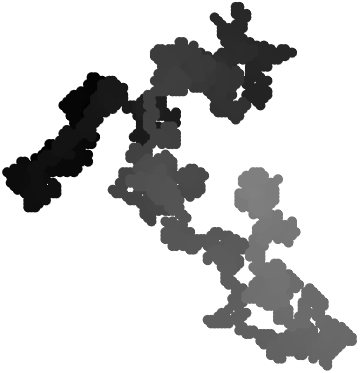},\includegraphics[scale=0.3]{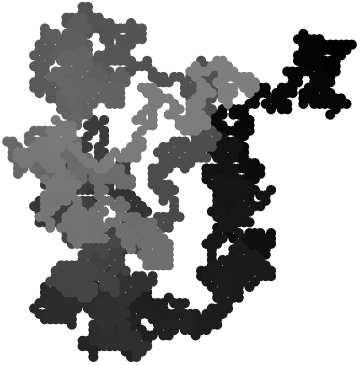},\includegraphics[scale=0.3]{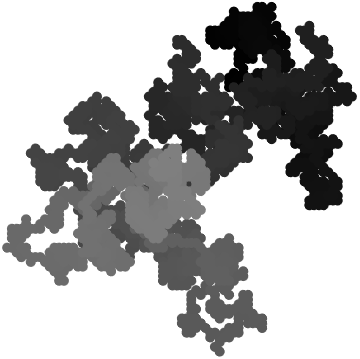}}
\end{center}
I believe that visualizing the numbers in a meaningful way is a very important project to pursue, we are only at the beginning of a great adventure!

To conclude, we can say that the Short Key Dream can also be realized in a different way: just whisper in your friend's ear  (assuming the necessary technical details have been  shared) "$\cos(5)$".

	\end{document}